%% file: elsearticle.tex
\documentclass[1p]{elsarticle}

\usepackage{lineno,hyperref}
\modulolinenumbers[5]

\journal{Computer Communications}

\usepackage{graphicx}
\usepackage[T1]{fontenc}
\usepackage[utf8]{inputenc}
\usepackage[english]{babel}
\usepackage[autostyle=true]{csquotes}
\usepackage{footnote}
\usepackage{textcomp}

\usepackage[caption=false]{subfig} 

\usepackage{algorithm}
\usepackage{algpseudocode}

\usepackage{tabularx}
\usepackage{booktabs} 
\usepackage{multirow}

\usepackage{xcolor}
\definecolor{burntorange}{rgb}{0.8, 0.33, 0.0}
\definecolor{chartreuse}{rgb}{0.87, 1.0, 0.0}

\usepackage{todonotes}

\usepackage{pifont}

\begin{document}

\begin{frontmatter}

\title{GeoBroker: Leveraging Geo-Contexts for IoT Data Distribution}

\author[1]{Jonathan Hasenburg\corref{cor1}}
\ead{jh@mcc.tu-berlin.de}

\author[2]{David Bermbach}
\ead{db@mcc.tu-berlin.de}

\address{Technische Universität Berlin \& ECDF, Einsteinufer 17, 10587 Berlin, Germany}

\cortext[cor1]{Corresponding author}

\begin{abstract}
In the Internet of Things, the relevance of data often depends on the geographic context of data producers and consumers.
Today's data distribution services, however, mostly focus on data content and not on geo-context, which could help to reduce the dissemination of excess data in many IoT scenarios.
In this paper, we propose to use the geo-context information associated with devices to control data distribution.
We define what geo-context dimensions exist and compare our definition with concepts from related work.

Furthermore, we designed GeoBroker, a data distribution service that uses the location of things, as well as geofences for messages and subscriptions, to control data distribution.
This way, we enable new IoT application scenarios while also increasing overall system efficiency for scenarios where geo-contexts matter by delivering only relevant messages.
We evaluate our approach based on a proof-of-concept prototype and several experiments.
\end{abstract}


\end{frontmatter}

\section{Introduction} \label{sec:introduction}
\input{01_introduction}

\section{Geo-Context Dimensions and Related Work} \label{sec:geo-context}
\input{sections/03_geocontext}

\section{GeoBroker} \label{sec:geobroker}
\input{sections/04_geobroker}

\section{Proof-of-concept Implementation} \label{sec:poc}
\input{sections/04b_implementation}

\section{Evaluation} \label{sec:evaluation}
\input{sections/05_evaluation}

\section{Discussion} \label{sec:discussion}
\input{sections/07_discussion}

\section{Conclusion} \label{sec:conclusion}
\input{sections/08_conclusion}

\section*{Acknowledgements}
\noindent This work was partially funded by the Deutsche Forschungsgemeinschaft (DFG, German Research Foundation) - 415899119.

\bibliography{bibfile}

\end{document}

%% file: 01_introduction.tex

Recent advances in mobile technologies and cyber-physical systems have led to a massive increase in data generation and distribution at the edge of the network.
A vision of the pervasive internet and the Internet of Things (IoT) is to make this data available across applications and devices~\cite{happ_meeting_2017} to enable new and better services for the society and industry.
For instance, efficient and intelligent communication between cars, bikes, and other road users could improve road safety~\cite{khelil_suitability_2014}.

Most data generated by IoT devices, however, is not relevant to the majority of other IoT devices and should not be distributed to reduce computational efforts and cope with bandwidth limitations.

Yet, today's data distribution services mostly focus on data content and not on the associated geo-context when deciding what data should be distributed to which clients.
Hence, they disregard information that is readily available in most IoT scenarios and should be used to improve the delivery precision.
For example, a car that aims to avoid traffic jams needs to process only data from roadside equipment and cars within its current neighborhood to determine an optimal route and velocity.
Therefore, from the perspective of data consumers, data originating outside an area of interest is not relevant and can be discarded.
A data producer, on the other hand, might already know that provided data is only relevant for data consumers in a specific area, and thus prevent others from receiving it.
E.g., only drivers in the immediate vicinity of a particular car might need to know its acceleration and deceleration profile which prevents data misuse.
Such scenarios are the reason why Bellavista et al.~\cite{bellavista_quality_2014} argue that geographical co-location should also be considered.
Other domains with applications in which the value of information depends on the location of data producers and consumers include the Internet-of-Vehicle \cite{murphy_context-aware_2007,sejin_shun_pub/sub-based_2016}, Smart Cities \cite{sanchez_smartsantander:_2014}, or Mobile Health \cite{nastic_serverless_2017}.

Furthermore, if a data producer trusts the location provided by a data consumer, the geo-context can be used as an alternative to credential-based authentication for data access control in some scenarios~\cite{trust}.
Consequently, a user-friendly middleware service that leverages geo-context information can not only help to limit the dissemination of excess data, but also enables new context-aware computing applications that have not been possible before.

Existing work has used spatial data for various reasons before, e.g.,~\cite{chapuis_scaling_2017,chapuis_horizontally_2017,chen_efficient_2003,murphy_context-aware_2007,guo_elaps:_2015,guo_location-aware_2015,herle_bridging_2016}.
{} 
Each of these papers, however, uses a different subset of geo-context information; neither paper covers the entire geo-context of data producers and consumers.
In addition, existing work does not evaluate the impact of using geo-context information on excess data dissemination.

In this paper, we propose a data distribution service that uses data content and the entire geo-context of clients to control data distribution.
Therefore, we make the following contributions:
\begin{itemize}
    \item We extend our previous geo-context model\footnote{This paper partially extends our previous work \cite{hasenburg_towards_2019}.} derived from related work (section~\ref{sec:geo-context}).
    \item We describe the design of GeoBroker, a data distribution service leveraging geo-contexts (section~\ref{sec:geobroker}).
    \item We describe a proof-of-concept implementation that we have made available as open source\footnote{\url{https://github.com/MoeweX/GeoBroker}} (section~\ref{sec:poc}).
    \item We evaluate the overhead of using geo-contexts, benchmark the performance of GeoBroker based on a realistic use case, and analyze the impact of using geo-contexts on excess data dissemination (section~\ref{sec:evaluation}).
\end{itemize}

Finally, we discuss important design decisions and limitations (section~\ref{sec:discussion}) before drawing a conclusion (section~\ref{sec:conclusion}).

%% file: sections/03_geocontext.tex

In this section, we first define and compare \enquote{content} and \enquote{context} (section~\ref{subsec:content_vs_context}).
Then, we present our geo-context model that builds upon and extends related work (section~\ref{subsec:geo-context}).
Finally, we discuss the related work (section~\ref{subsec:rw}).

\subsection{Content vs Context} \label{subsec:content_vs_context}

In the following, we explain and compare \enquote{content} and \enquote{context} with the help of a topic-based pub/sub system.
In such a system, publishers are the data producers and subscribers are the data consumers.
Subscribers define which \textbf{content} they are interested in by subscribing to topics, e.g., when a subscriber creates a subscription to the topic \textit{sensor/temperature}, it receives temperature sensor measurements published to the same topic.

Dey defines \textbf{context} as \enquote{any information that can be used to characterize the situation of an entity} \cite{dey_understanding_2001}.
Thus, the context of an IoT device comprises many items such as other nearby devices or the type of used power source.
In this paper, we only focus on the geo-context which we consider to comprise (1) the location of the device and (2) special areas that are of interest/relevance to the device.

So why is it necessary to distinguish between content and geo-context?
Both producers or consumers may have moved in between sending and receiving two data items.
This, however, is not reflected in the content-related interests (e.g., the subscription) but affects the context-related interests.
Hence, location information is not related to content.

Distinguishing content and geo-context information also has many practical benefits.
For example, while it is possible to encode some geo-context information in topics, this requires clients to agree on such a structure and leads to very complicated and bloated topic trees.
E.g., one could agree that the first topic level is always the country and the second topic level is always the city to which a given message refers.
Then, the topic \textit{france/paris/sensor/temperature} would refer to all temperature sensors in Paris, while the topic \textit{germany/berlin/sensor/ tem\-pe\-ra\-ture} would refer to all temperature sensors in Berlin.
Besides the disadvantages mentioned above, this approach is very coarse-grained and it is not possible to distinguish between the location of a device and its area of interest.

\subsection{Four Dimensions of Geo-Context} \label{subsec:geo-context}

Previous work has already proposed to use geo-context information for more advanced control of data distribution.
Their focus, however, has not been on developing a general view on IoT device geo-contexts.
Instead, the authors typically designed a system for a particular use case in which location-based data needs to be processed; thus, they do not consider all geo-context dimensions but rather only those relevant to their specific use case.
Based on this related work, we derived a generally applicable definition of the entire geo-context.
In the following, we first present this definition so that we can use the terminology to discuss which dimensions have been considered in related work (section~\ref{subsec:rw}).

We identified four geo-context dimensions.
Both data producers and data consumers have a geographic location (\textbf{producer location} and \textbf{consumer location}), which consists of a latitude and a longitude value\footnote{For stationary clients, such locations may already be known and provided as a parameter when starting the client. In all other cases, the location can be determined using a GPS sensor or approaches such as WiFi trilateration~\cite{rusli_improved_2016}.}.
Beyond this, data producers and data consumers each have an area of interest; we propose to use geofences
\footnote{A Geofence is a virtual fence surrounding a defined geographical area. As a usage example, Reclus and Drouard describe a scenario in which such fences are used to notify factory workers about approaching trucks~\cite{reclus_geofencing_2009}.
For our purposes, a geofence can have arbitrary shapes and may comprise non-adjacent subareas, e.g., Germany and Italy.}
to describe these areas; these can, for instance, be specified in the \texttt{Well-Known Text}~\cite{wkt} format.
The \textbf{consumer geofence} ensures that received data originates in an area of interest, i.e., producer locations are inside the consumer geofence.
The \textbf{producer geofence}, on the other hand, ensures that only data consumers present in a specific area receive data, i.e., consumer locations are inside the producer geofence.

As in the case of data content, producers and consumers can have multiple geo-contexts.
For example, in a topic-based pub/sub system, a subscriber (consumer) can create individual subscriptions for different topics.
Thus, when also using geo-contexts, subscribers might specify a geofence per subscription.
Likewise, publishers might specify a geofence for every message.

\begin{figure}[ht]
    \centering
    \includegraphics[width=0.85\textwidth]{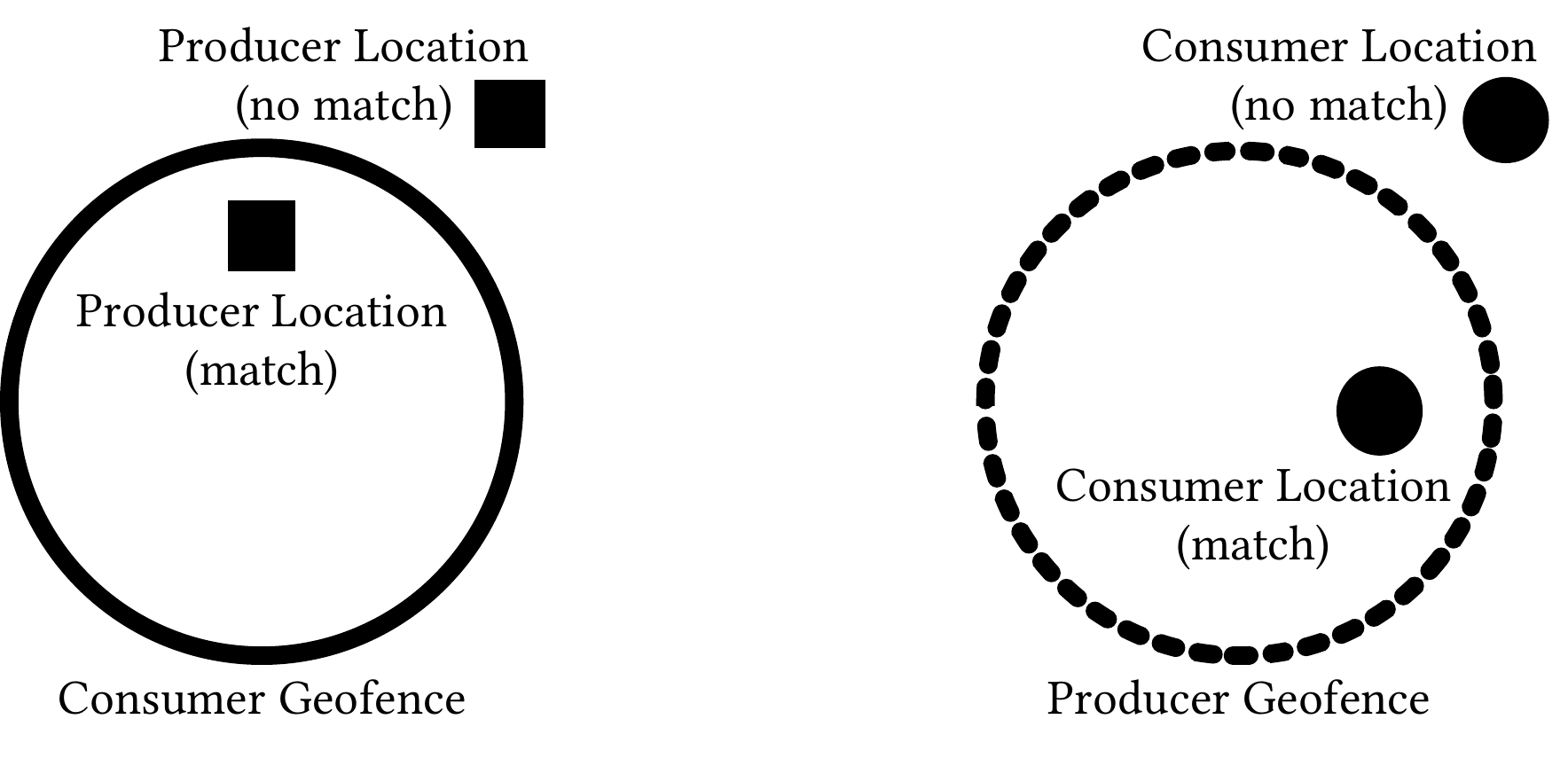}
    \caption{Consumer GeoCheck (Left) and Producer GeoCheck (Right)}
    \label{fig:geofence_abstract_example}
\end{figure}

For bringing geofences and locations together, two checks are necessary to decide whether data from a given producer should be sent to a given consumer (figure~\ref{fig:geofence_abstract_example}) --
first, from the consumer's perspective with the help of the consumer geofence and the producer location (consumer GeoCheck) and, second, from the producer's perspective with the help of the producer geofence and the consumer location (producer GeoCheck).

A data consumer can limit data distribution if it already knows that only data from a certain area is relevant and wishes to avoid being overloaded by excess data.
Data producers, on the other hand, might have two motivations for limiting distribution.
First, as a form of access control if the location can be sufficiently trusted, e.g., only visitors of a building should get access to its smart home data.
Second, to use domain knowledge that is not available to the consumers. For example, if a data producer sends out wireless emergency alerts\footnote{\footnote{Wireless Emergency Alerts - \url{https://www.fcc.gov/consumers/guides/wireless-emergency-alerts-wea}}.}, only the producer knows which area is affected by the corresponding event.
For a more detailed discussion of use cases, we refer to our previous work \cite{hasenburg_towards_2019}.
There are, however, also scenarios in which a consumer or producer does not wish to limit data distribution.
Then, depending on the implementation, they can either supply no geofence or a geofence that comprises all clients.

\subsection{Related Work} \label{subsec:rw}

Spatial information is used for various purposes in many research areas, e.g., such information can be used to optimize replica placement~\cite{yu_location-aware_2015}.
For our related work discussion, however, we focus on more closely related approaches, i.e., approaches that enhance data distribution through geo-context information.
At the end of this section, we also give an overview of related work on general (IoT) data distribution approaches and techniques.

Table~\ref{tbl:rw_table} summarizes which of the four geo-context dimensions are considered by related work.
Note, that we use our geo-context dimension terminology from section~\ref{subsec:geo-context} for the following discussion.

\begin{table}[ht]
\centering
\begin{tabular}{@{}ccccccc@{}}
\toprule
\multicolumn{1}{l}{}                                  &                      & \multicolumn{2}{c}{Location} &                      & \multicolumn{2}{c}{Geofence} \\ \midrule
RW                                                    & \phantom{s}          & Consumer      & Producer     & \phantom{s}          & Consumer     & Producer      \\ \midrule
\cite{chen_efficient_2003}                            &                      & Yes           & No           &                      & No           & Yes           \\
\cite{guo_elaps:_2015,guo_location-aware_2015}        &                      & No            & Yes          &                      & Yes          & No            \\
\cite{chow_towards_2010}                              &                      & No            & Yes          &                      & Yes          & No            \\
\cite{bryce_mqtt-g:_2018}                             &                      & No            & Yes          &                      & Yes          & No            \\
\cite{li_location-aware_2013}                         &                      & No            & Yes          &                      & Yes          & No            \\
\cite{wang_ap-tree:_2015}                             &                      & No            & Yes          &                      & Yes          & Partially     \\
\cite{chapuis_scaling_2017,chapuis_horizontally_2017} &                      & Partially     & Partially    &                      & Yes          & Yes           \\
\cite{murphy_context-aware_2007}                      &                      & Partially     & Partially    &                      & Yes          & Yes           \\
\cite{herle_bridging_2016,herle_enhancing_2018}       &                      & Yes           & Yes          &                      & Yes          & Yes           \\ \bottomrule
\end{tabular}
\caption{An Overview of the Geo-Context Dimensions Considered by Related Work}
\label{tbl:rw_table}
\end{table}

Chen et al.~\cite{chen_efficient_2003} propose a spatial middleware service that delivers messages to clients when they enter \enquote{zones} defined by data producers, i.e., producer geofences.
While this allows data producers to control data distribution based on areas they consider as relevant, data consumers cannot control data distribution with consumer geofences.

The authors of~\cite{guo_elaps:_2015,guo_location-aware_2015,chow_towards_2010,bryce_mqtt-g:_2018,li_location-aware_2013} consider the consumer geofence and the producer location, i.e., the geo-context dimensions needed for the consumer GeoCheck.
However, neither group of authors lets data producers control the matching of messages based on producer geofences and consumer locations (producer GeoCheck).
Guo et al.~\cite{guo_elaps:_2015,guo_location-aware_2015} propose a location-aware pub/sub service that delivers messages based on consumer geofences attached to subscriptions.
Chow et al.~\cite{chow_towards_2010} present GeoSocialDB, a system that provides three location-based services for social networks: a news feed, news ranking, and recommendations.
Each of these services can be queried to retrieve all data of data producers that are located (producer location) in an area defined by a data consumer (consumer geofence).
Bryce et al.~\cite{bryce_mqtt-g:_2018} propose MQTT-G, an extension of the MQTT protocol with Geolocation.
While subscribers can define consumer geofences to control message distribution, their geofences are only created once per subscriber rather than for individual subscriptions.
Li et al.~\cite{li_location-aware_2013} propose to use an R-Tree index structure to efficiently identify which data producers are located in areas defined by data consumers.
Again, these groups of authors only look at geo-context from one perspective, so their approaches do not work with producer geofences and consumer locations.

Wang et. al~\cite{wang_ap-tree:_2015} propose the AP-Tree to efficiently support location-aware pub/sub.
While they only discuss the producer location and consumer geofence in their paper, they state that using a producer geofence is also possible with their approach.
With this geofence alone, however, they cannot run the producer GeoCheck as the consumer location is required as well.

Chapuis et al.~\cite{chapuis_scaling_2017,chapuis_horizontally_2017} propose a horizontally scalable pub/sub architecture that supports matching based on a circular geofence around publishers and around subscribers.
Hence, their message matching does not consider client locations independent of geofences.
Furthermore, their geofences can only be circular while GeoBroker supports arbitrary geofence sizes.

Frey and Roman~\cite{murphy_context-aware_2007} also propose a protocol to bring context to pub/sub systems.
They allow publishers to define a \enquote{context of relevance}, and subscribers to define a \enquote{context of interest}.
When both contexts overlap, a message is delivered to the subscriber.
While their context definition is very general, it can also be used for geo-context information, i.e., the (1) location of a device and (2) producer/consumer geofences.
However, they mix geofences and locations, so if a client moves it needs to update its subscriptions even if the consumer geofence did not change.

Herle et al.~\cite{herle_bridging_2016,herle_enhancing_2018} propose to extend the MQTT protocol so that messages can be matched based on locations and geofences appended to published messages and subscriptions.
Each subscription or published message, however, can have either a geofence \underline{or} a location.
Thus, it is only possible to do one of the two GeoChecks, but never both.

Beyond the geo-context related work, there is also related work on (IoT) data distribution approaches and techniques; some of these also consider a subset of the above introduced geo-context dimensions.
We see the most closely related approaches in the area of pub/sub, e.g., in IoT~\cite{banno_designing_2015,chen_coss:_2015,rausch_emma:_2018,sun_low-delay_2013,teranishi_scalable_2015} but also for applications ranging from enterprise to web computing~\cite{barazzutti_streamhub:_2013,moro_comparative_2007,cugola_introducing_2005,goos_supporting_2003,gascon-samson_dynamoth:_2015,gascon-samson_multipub:_2017,pietzuch_hermes:_2002,sasu_tarkoma_publish/subscribe_2012,sharma_wormhole:_2015}.
In addition, there is also non-pub/sub related work on data distribution in general, e.g., \cite{g._werner-allen_deploying_2006,paluska_footloose:_2003,goos_strategies_2000}; neither of these explicitly captures geo-context information.

%% file: sections/04_geobroker.tex

In this section, we describe the design of GeoBroker, our data distribution service that leverages the full geo-context of data producers and data consumers to reduce excess data dissemination and facilitate the development of new, pervasive IoT applications.
GeoBroker offers a similar functionality and operational behavior as popular services already used today, while adding the capabilities necessary for the two GeoChecks that we have introduced in section \ref{subsec:geo-context}.

Widely used IoT data distribution services such as AWS IoT\footnote{\url{https://aws.amazon.com/iot/}} or Google Cloud IoT\footnote{\url{https://cloud.google.com/solutions/iot/}} build upon pub/sub to provide asynchronous, loosely coupled communication between publishers (the data producers), and subscribers (the data consumers), that do not even have to know each other~\cite{paridel_middleware_2010}.
Such services often rely on topic-based pub/sub because it is lightweight, payload agnostic, and works well with IoT devices that operate in constrained environments.

Therefore, GeoBroker is designed as a topic-based pub/sub system as well.
Note, however, that our approach could also be applied to content-based pub/sub systems if required\footnote{Bellavista et al. see topic-based pub/sub as a particular instance of content-based pub/sub~\cite{bellavista_quality_2014}, but rather than using the entire message for matching only parts of it are used (the topic) which makes it more lightweight.}.
Furthermore, the concept of using the geo-context for a more precise data distribution could also be used for data distribution that is not based on pub/sub, e.g., for replica placement decisions in a geo-distributed storage system.

GeoBroker provides basic pub/sub broker functionality, i.e., clients can connect, subscribe, and publish messages (see section~\ref{subsec:operational_behaviour}).
However, we extended the matching of published messages and subscriptions to consider geo-contexts as well (see section~\ref{subsec:matching_process}).
To do this efficiently, we designed a data structure that indexes subscriptions (see sections~\ref{subsec:indexing_structure} and \ref{subsec:updating_subscriptions}).

\subsection{GeoBroker Functionality} \label{subsec:operational_behaviour}

GeoBroker is topic-based and has an operational behavior similar to that of many other pub/sub systems; we extended this behavior where necessary.
To ease integration into real systems beyond research prototypes, we use the same message types as MQTT~v5.0~\cite{mqtt_v5}, a widely used pub/sub protocol, but piggyback geo-context information on top of them; so we also use a different message format.

As in a typical MQTT system, clients first connect to GeoBroker and create a session.
These sessions expire after a certain time so that clients periodically send keep-alive (ping) messages to GeoBroker.
In ``connect'' and ``ping'' messages, clients also includes their current location which is stored by GeoBroker; this applies to clients acting as data consumer or as data producer.
Similarly to most other messages, GeoBroker acknowledges connect and ping messages so that lost messages can be detected.

Comparable to MQTT, clients can act as data consumer and data producer at the same time.
Data consumers with an active session can create and delete topic subscriptions.
In GeoBroker, however, they can also update the consumer geofence of each subscription.
Whenever a data producer publishes a message, GeoBroker matches the topic and geo-context information of the published message with the information of its managed subscriptions and distributes the message accordingly to data consumers.

Sessions terminate when the respective client sends a disconnect message or after a time-out.
Terminating a client session also suspends all active subscriptions.

An essential characteristic of broker-based pub/sub systems is that clients are completely decoupled:
Data producers can publish messages to GeoBroker without having to worry whether any data consumers are connected, ready to receive messages, or crashed as this is handled entirely by GeoBroker.
In addition, clients can act as data consumers and producers at the same time.

\subsection{Message Matching} \label{subsec:matching_process}

GeoBroker extends the vanilla message matching and also considers geo-context information rather than content information (in the form of topics) only.
For each published message that GeoBroker receives, GeoBroker runs the following checks to determine to which data consumers it should distribute a data producer's message:
\begin{enumerate}
    \item ContentCheck: checks whether the subscription topic matches the message topic.
    \item Consumer GeoCheck: checks whether the consumer geofence contains the producer location.
    \item Producer GeoCheck: checks whether the producer geofence of the message contains the consumer location.
\end{enumerate}
Then, GeoBroker delivers the message to the data consumers who passed all three checks.
We explicitly decided on this order as the ContentCheck requires less computation than the two GeoChecks.
In corner cases, where clients are spread across a large area, however, running the GeoChecks first might be more efficient.
Regarding the ordering of consumer and producer GeoCheck, we run the consumer GeoCheck first as consumer geofences can be efficiently stored and indexed in advance whereas the producer geofence is not known until the respective message arrives at GeoBroker.
There are some scenarios that do not require all four geo-context dimensions, so the related GeoCheck can be skipped.

\subsubsection{ContentCheck} \label{subsec:topic-matching}

GeoBroker does the same ContentCheck as MQTT-based systems, i.e., it matches the topics of subscriptions and published messages~\cite{mqtt_v5}:
Topics are identified by their names, which may consist of multiple levels separated by \enquote{/}.
For example, if a data consumer creates a subscription for the topic \textit{a/b}, it will receive all messages published to the topic \textit{a/b}, but no other messages, e.g., published to topic \textit{a/c}.
Besides such fixed topics, data consumers can also use special wildcard characters to subscribe to multiple topics at once.
A wildcard is either valid at a single topic level (\enquote{+}) or at multiple topic levels (\enquote{\#}).

\subsubsection{Consumer GeoCheck} \label{subsec:subscriber-geo-matching}

The consumer GeoCheck is only run on subscriptions that passed the ContentCheck.
The check should be carried out by an efficient spatial indexing structure.
When a data consumer creates or updates a subscription, the consumer geofences (for the given topic) has to be stored in the indexing structure.
For the consumer GeoCheck with a given topic, the subscription indexing structure must return all subscriptions with a consumer geofence that contains the producer location already known to GeoBroker\footnote{As explained in section~\ref{subsec:operational_behaviour}, the data producer sets and updates its location with connect and ping messages.}.
We describe the design of our data structure in section~\ref{subsec:indexing_structure}.

\subsubsection{Producer GeoCheck} \label{subsec:publisher-geo-matching}

The producer GeoCheck is only run on subscriptions that passed the ContentCheck and the consumer GeoCheck.
In this final step, GeoBroker checks whether the producer geofence of the message contains the corresponding consumer locations.
If so, it delivers the message to the respective data consumer.

\subsection{Subscription Indexing Structure} \label{subsec:indexing_structure}

All information necessary for the ContentCheck and consumer GeoCheck is available before GeoBroker processes a message.
Thus, GeoBroker can store subscription related information in a data indexing structure for efficient retrieval.
Note, that the producer geofence is part of the published message and is thus not available beforehand, so the producer GeoCheck cannot be supported by such a data structure.

Approaches for spatial-keyword matching exist already today, e.g., Wang et al. proposed the AP-Tree~\cite{wang_ap-tree:_2015} and showed that it is more efficient than other solutions.
With spatial-keyword matching, however, ContentCheck and consumer GeoCheck information are stored in the same data structure, so it is non-trivial/challenging to change the type from topic-based to content-based and vice versa.
Therefore, we designed our own subscription indexing structure that
\begin{itemize}
    \item is capable of first running the ContentCheck before continuing to the consumer GeoCheck,
    \item has a low updating overhead as data consumers might be mobile and use consumer geofences that move with them,
    \item supports multi-threading.
\end{itemize}
The main idea of our approach is using a standard indexing structure for the ContentCheck and embedding a second data structure inside to efficiently run the consumer GeoCheck.


In the case of GeoBroker, the ContentCheck is done based on MQTT topics.
Popular MQTT brokers such as mosquitto\footnote{https://mosquitto.org} or moquette\footnote{https://github.com/andsel/moquette} use a directed rooted tree to efficiently match topics.
Therefore, we use a similar \textbf{topic tree} structure for the ContentCheck which stores topic levels in tree nodes.
\begin{figure}[ht]
    \centering
    \includegraphics[width=\textwidth]{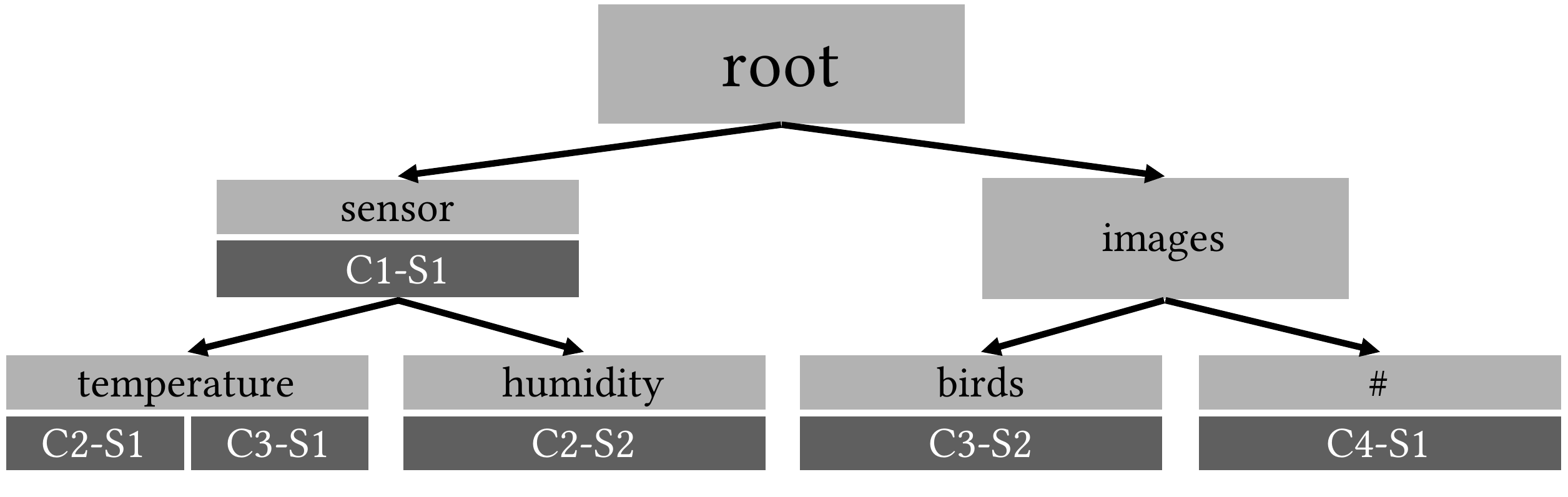}
    \caption{Topics are Stored in a Directed Rooted Tree}
    \label{fig:topic_tree}
\end{figure}
Figure~\ref{fig:topic_tree} shows an example of such a topic tree in which data consumers (C1 -- C4) have created various subscriptions for specific topics (e.g., \textit{sensor/temperature}) and a wildcard topic (\textit{images/\#})\footnote{More information on topic filters can be found in the MQTT v5.0 OASIS Standard~\cite{mqtt_v5}}.
When GeoBroker receives a published message, it traverses the tree until it finds the node that stores the matching topic and thus all matching subscriptions.
Note, that with wildcards it is possible to find multiple nodes.

\begin{figure}[ht]
    \centering
    \includegraphics[width=\textwidth]{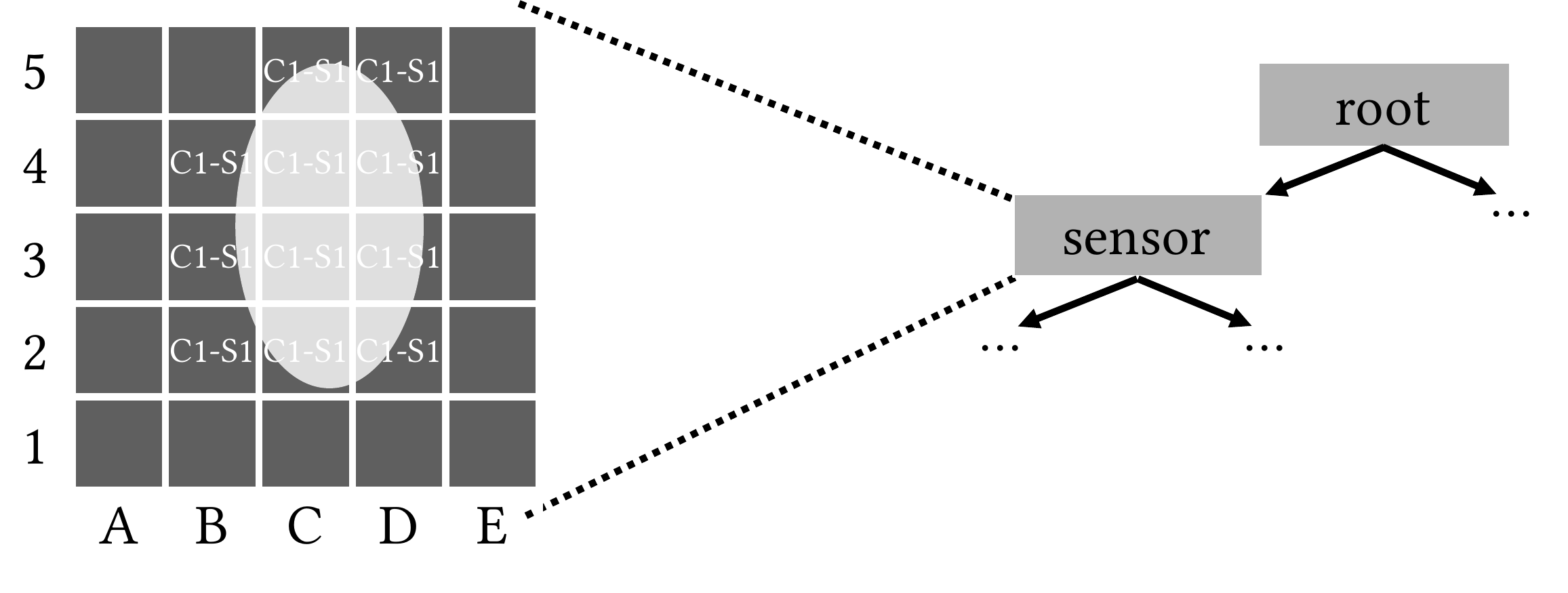}
    \caption{Nodes in the Topic Tree Contain a Raster as Embedded Spatial-Indexing Data Structure}
    \label{fig:raster}
\end{figure}

Each tree node of the topic tree contains a \textbf{raster} as embedded spatial-indexing data structure (see also figure~\ref{fig:raster}) that allows GeoBroker to efficiently identify the subscriptions that contain a given producer location -- in particular, the raster helps to reduce the number of ``contains'' operations for geofences.

A raster stores all corresponding subscriptions in a 2D data structure that divides the available geographic space (e.g., the surface of the earth) into rectangular areas (\textbf{raster fields}).
Raster fields can be uniquely identified and accessed via the coordinate of their respective southwest corner.
Furthermore, raster fields do not overlap, directly border each other, and exactly one has its southwest corner at the point of origin (0\textdegree/0\textdegree).

Each raster field contains a list of all subscriptions that have an intersecting consumer geofence; see figure~\ref{fig:raster} for an example showing a subscription created by data consumer C1 that targets the topic \textit{sensor} and has an almost circular geofence.
The idea behind this is to reduce the number of ``contains'' checks that are required to identify which consumer geofences contain a given producer location as this check becomes increasingly compute-intensive with more complex shapes.
With the raster, only the subscriptions that are referenced in the same raster field as the one containing the producer location need to be checked for intersection.

Smaller raster fields mean that fewer consumer geofences need to be che\-cked.
However, this makes subscription updates more costly and increases the overall size of our index data structure as each subscription reference needs to be added to/removed from more raster fields.
For this obvious tradeoff, the raster field size provides the tuning knob to balance costs of message matching and subscription updates.
In addition, the \enquote{optimal} raster field size also depends on the average geofence size and shape.
To control the raster field resolution, we added a parameter called granularity.
We then defined the side length of each raster field as \textit{1\textdegree /granularity}\footnote{We chose degree rather than meter as measurement unit since the earth has a spherical surface and our raster fields remain quasi-rectangular. However, choosing degree has the downside of raster fields becoming smaller when moving away from the equator in terms of their real size when measured in meters.}, so when a user increases the granularity, the raster fields become smaller.

Instead of our own raster approach, we could have used another spatial indexing structure such as an R-Tree~\cite{Hadjieleftheriou2017,lee_supporting_2003} or a B-Tree, which stores spatial regions encoded as bit strings~\cite{skovsgaard_top-k_2014}.
In fact, we tried both approaches before coming up with our raster-based design but faced performance issues.
We did an experiment with 110k geofence operations (35k adds, 25k updates, 50k gets) executed in a single-thread.
Our R-Tree implementation needed about 72 seconds, most time was spent on traversing the tree due to the necessary bounding box checks.
Our B-Tree implementation needed about 232 seconds.
Here, most time was spent on identifying potential keys and removing false positive subscriptions.
For the same setup, our raster-based solution completed all operations in about 1.6 seconds.

At this point, we would like to emphasize that optimizing the subscription indexing structure is not the focus of this paper.
However, first micro-benchmark results -- as indicated above -- show that our proposed indexing structure has a high performance.
In future work, we plan to further explore and compare different data structures to identify the best possible solution.

\subsection{Updating Subscriptions} \label{subsec:updating_subscriptions}

Each time a data consumer subscribes or unsubscribes, the topic tree and raster need to be updated.
In the following, we discuss how this is done when a new subscription is added as the steps for updating or removing a subscription are almost identical.

To create a new subscription, GeoBroker first traverses the topic tree until it finds the node which corresponds to the subscription topic.
Then, GeoBroker determines the raster fields that intersect with the consumer geofence as the subscription has to be added to these fields.
It is very compute-intensive to identify these out of all the raster fields stored in the raster, so GeoBroker only checks the ones that intersect with the area inside the outer bounding box of the geofence. Identifying these is inexpensive when done with algorithm~\ref{alg:raster_rectangle_algorithm} using the raster field keys\footnote{The computational effort scales linearly with the number of raster fields inside the bounding box.}. For efficient removal of old subscriptions, GeoBroker can either cache the consumer geofences or data consumers can provide the old geofence as part of their request.

\begin{algorithm}[ht]
\caption{Updating Subscriptions: Identify Raster Fields That Intersect With the Area Inside a Subscription Geofence Bounding Box}
\label{alg:raster_rectangle_algorithm}
\begin{algorithmic}
    \Function{calculateKey}{location}
    \State lat = floor(location.lat * granularity) / granularity
    \State lon = floor(location.lon * granularity) / granularity
    \State \Return (lat/lon)
    \EndFunction
    \State
    \Function{main}{}
    \State swInd = calculateKey(southWestBoundingBoxCorner)
    \State neInd = calculateKey(northEastBoundingBoxCorner)
    \For{lat = swInd.lat \textbf{To} neInd.lat \textbf{Step} 1 / granularity}
        \For{lon = swInd.lon \textbf{To} neInd.lon \textbf{Step} 1 / granularity}
            \State results.add(raster field with key (lat/lon))
        \EndFor
    \EndFor
    \State \Return results
    \EndFunction
\end{algorithmic}
\end{algorithm}

\begin{figure}[ht]
    \centering
    \includegraphics[width=.7\textwidth]{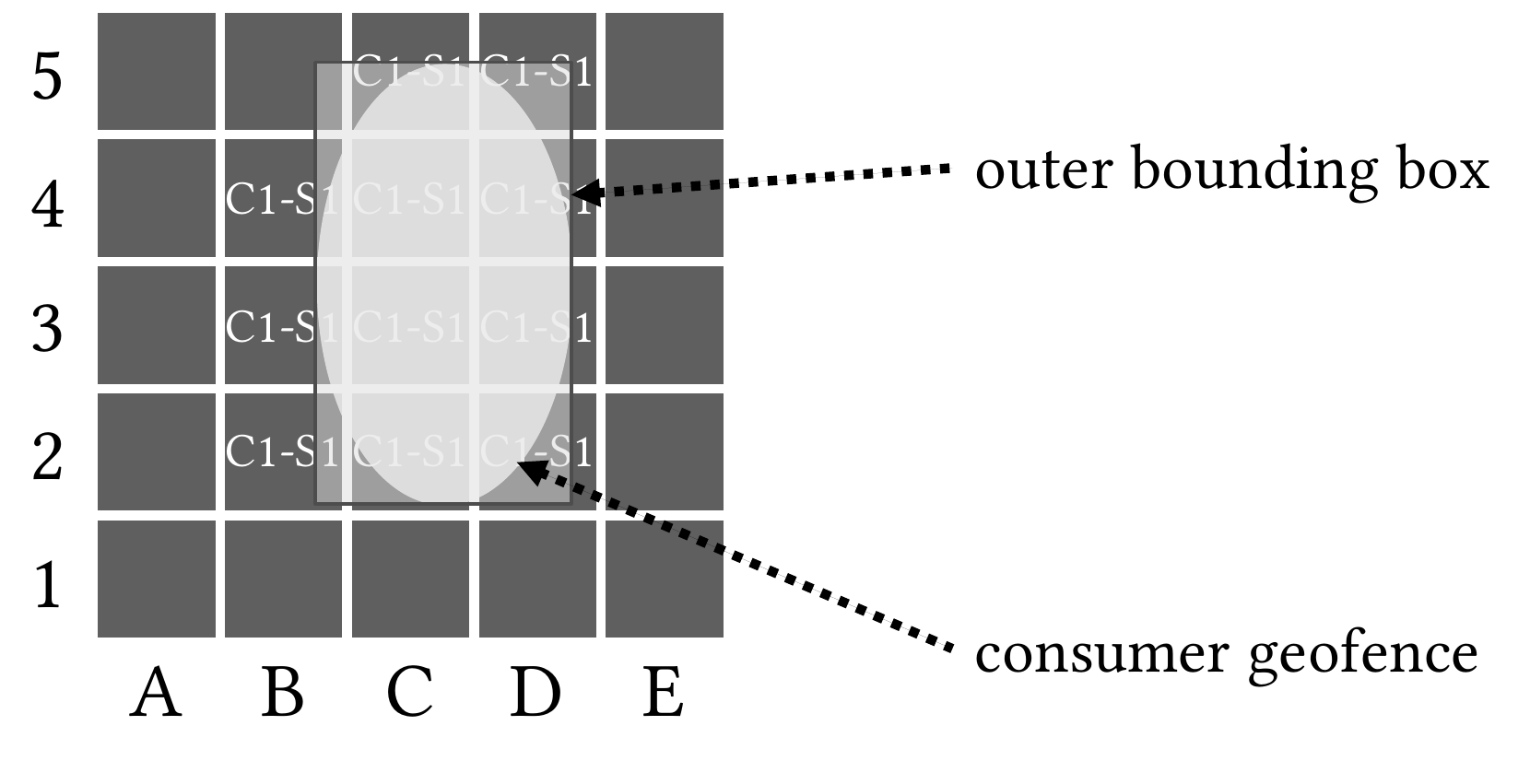}
    \caption{It is only Necessary to Check the Raster Fields inside the Geofence's Outer Bounding Box for Intersection with the Geofence when Updating Subscriptions}
    \label{fig:bounding_box_check}
\end{figure}

As an example, consider figure~\ref{fig:bounding_box_check} in which the raster fields B2 to B5, C2 to C5, and D2 to D5 are intersecting with the outer bounding box.
Note, that B5 is a false positive in this case, as it intersects with the outer bounding box but not with the geofence itself.
Thus, in a second step, it is necessary to additionally check each identified raster field for intersection with the geofence.

In theory, it is possible to omit this final intersection check, as false positives caused by wrongly added subscriptions to raster fields are filtered during the consumer GeoCheck (section~\ref{subsec:subscriber-geo-matching}).
This, however, only makes sense if a workload is very update-heavy and shapes have circular or rectangular patterns as otherwise too many false positives need to be filtered by the consumer GeoCheck that is run for every message that shall be published.

\subsection*{Summary}

In this section, we proposed the design of GeoBroker, our data distribution service leveraging geo-contexts.
GeoBroker offers a similar functionality and operational behavior as popular data distribution services already used today.
We extended, however, the message matching with two GeoChecks, one from the consumer and one from the producer perspective, in addition to the widely done ContentCheck to reduce excess data dissemination.
GeoBroker uses a novel, efficient subscription indexing structure that we specifically designed for the message matching with content and geo-context information.
With GeoBroker, each of the scenarios and examples that we have introduced throughout this paper can be implemented through a very basic API that relies on data consumers creating subscriptions and data producers publishing messages; we adapted both types of operations to support content and geo-context information.
GeoBroker is a very general solution, i.e., the service could be used by various applications for different purposes simultaneously.

%% file: sections/04b_implementation.tex

As a proof-of-concept, we implemented the data distribution service GeoBroker and a client in Java 8 and Kotlin with the functionality described in section~\ref{subsec:operational_behaviour}.

For the communication between GeoBroker and clients, we use the Java version of ZeroMQ\footnote{https://github.com/zeromq/jeromq}.
ZeroMQ is a networking library that builds on top of a high-speed asynchronous I/O engine~\cite[p. xiii ff.]{hintjens_zeromq:_2013}.
Its sockets can communicate in-process, inter-process, via TCP, and multicast, so it is not just a networking library but can also be used as a concurrency framework.
As ZeroMQ manages connections, a single socket can be used to handle thousands of clients.

In contrast to vanilla MQTT messages, which are encoded as defined by the MQTT v5.0 protocol, we serialize Java messages with Kryo\footnote{Kryo - \url{https://github.com/EsotericSoftware/kryo}} before handing them over to ZeroMQ.
So while our message types are similar to the ones of MQTT (e.g., we have a CONNECT message to establish a connection between clients and GeoBroker and a CONNACK message to acknowledge a connection), the messages themselves look different.
We chose that approach as it allows us to easily enhance the messages with additional information, while also not forcing us to implement all MQTT messaging features.
For example, the PINREQ message, which is used by clients to reset their session timers, does not support carrying a payload originally; for our approach, however, we use the ping functionality to update client locations, so appending a payload to this type of message is necessary.

\begin{figure}[ht]
    \centering
    \includegraphics[width=\textwidth]{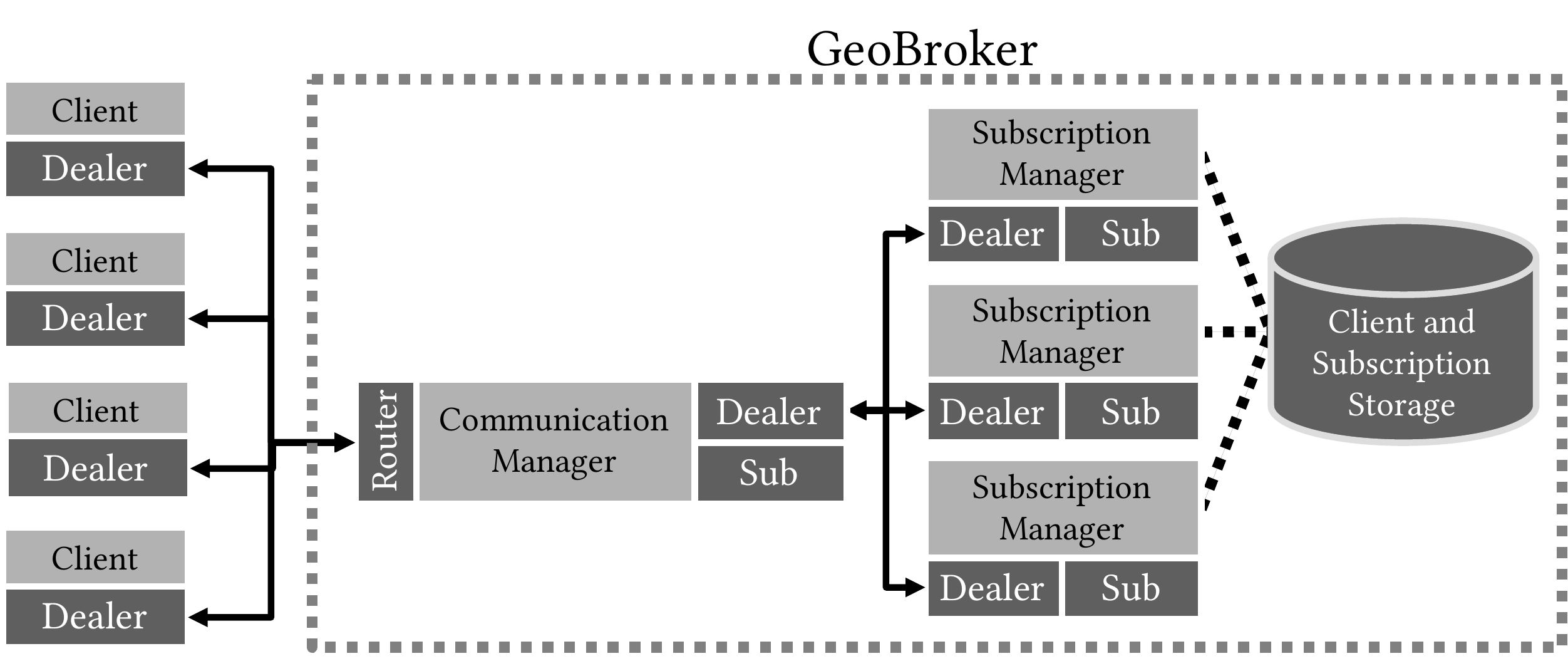}
    \caption{GeoBroker uses ZeroMQ for Internal and External Communication}
    \label{fig:zeromq}
\end{figure}

As ZeroMQ can be used as a concurrency framework, we also use ZeroMQ for GeoBroker's internal communication between threads to make it scale well.
Figure~\ref{fig:zeromq} shows a simplified version of the GeoBroker architecture.
Clients use a ZeroMQ dealer sockets to connect to the ZeroMQ router socket of the GeoBroker communication manager (this provides asynchronous communication between both parties~\cite[p. 88]{hintjens_zeromq:_2013}).
Internally, GeoBroker uses an arbitrary number of subscription managers which each runs in a separate thread.
These subscription managers use the client and subscription storage to manage connected data producer as well as data consumers and their active subscriptions.
As the storage implements our subscription indexing structures for content and geo-context information (section~\ref{subsec:indexing_structure}), it can efficiently retrieve the information needed for the three message matching checks.
The subscription manager uses a dealer socket to connect to the dealer socket of the communication manager;
this socket type combination gives us asynchronous communication between both parties as well; however, when the communication manager signals that a data producer's message is available for processing, only a single subscription manager will receive it.
The communication manager itself has virtually no load as it only forwards messages to the subscription managers.

GeoBrokers internal components also use a Sub socket to receive broadcasted instructions such as \enquote{shut down}\footnote{To do that via such a Sub socket is recommended~\cite[p. 57f.]{hintjens_zeromq:_2013}.}.
In the figure, we exclude the broadcasting components to improve readability.

%% file: sections/05_evaluation.tex

In this section, we present the evaluation of GeoBroker.
We used the Geolife dataset~\cite{zheng_geolife:_2010} to generate realistic workloads (section~\ref{subsec:workload}) for the evaluation of the GeoCheck overhead (section~\ref{subsec:geomatching_overhead}) and for a use case evaluation (section~\ref{subsec:use_case}).

\subsection{Generating a Realistic Workload} \label{subsec:workload}

To evaluate the performance and the overhead of GeoBroker, we need a dataset with spatial information.
We picked the Geolife V1.3 data set~\cite{zheng_geolife:_2010} which contains 18,670 GPS trajectories collected over five years. More than 90\% of the trajectories contain one entry every one to five seconds, and each entry comprises a timestamp and the current location of the user.

Based on the data set, we implemented two types of clients for different workloads, \emph{TravelClient} and \emph{TeleportingClient}.
The ClientManager (see below) is responsible for starting both types of client. When a client is started, it is initialized with one trajectory from the data set and keeps \enquote{traveling} along the corresponding route of locations until it is shut down.
When a client arrives at a location, it executes several pre-defined operations; what these operations are depends on the desired workload type, e.g., updating a subscription or publishing a message.
The TravelClient uses the timestamps to determine how much time it takes to arrive at the next location, the TeleportingClient ignores the timestamps and processes the trajectory as fast as possible.
When the last location of a trajectory has been reached, both types of client immediately jump back to the first location of their trajectory and start to travel again.
Each client runs in its own thread, thus, implementing a closed workload model~\cite{book_bermbach_cloud_service_benchmarking}.

For the evaluation, three operations are of particular interest:
First, updating the current client location.
Second, updating subscriptions for a given client which includes removing the old one first, if a client has already created one for a certain topic.
Third, publishing a message.

\begin{figure}[ht]
    \centering
    \includegraphics[width=0.7\columnwidth]{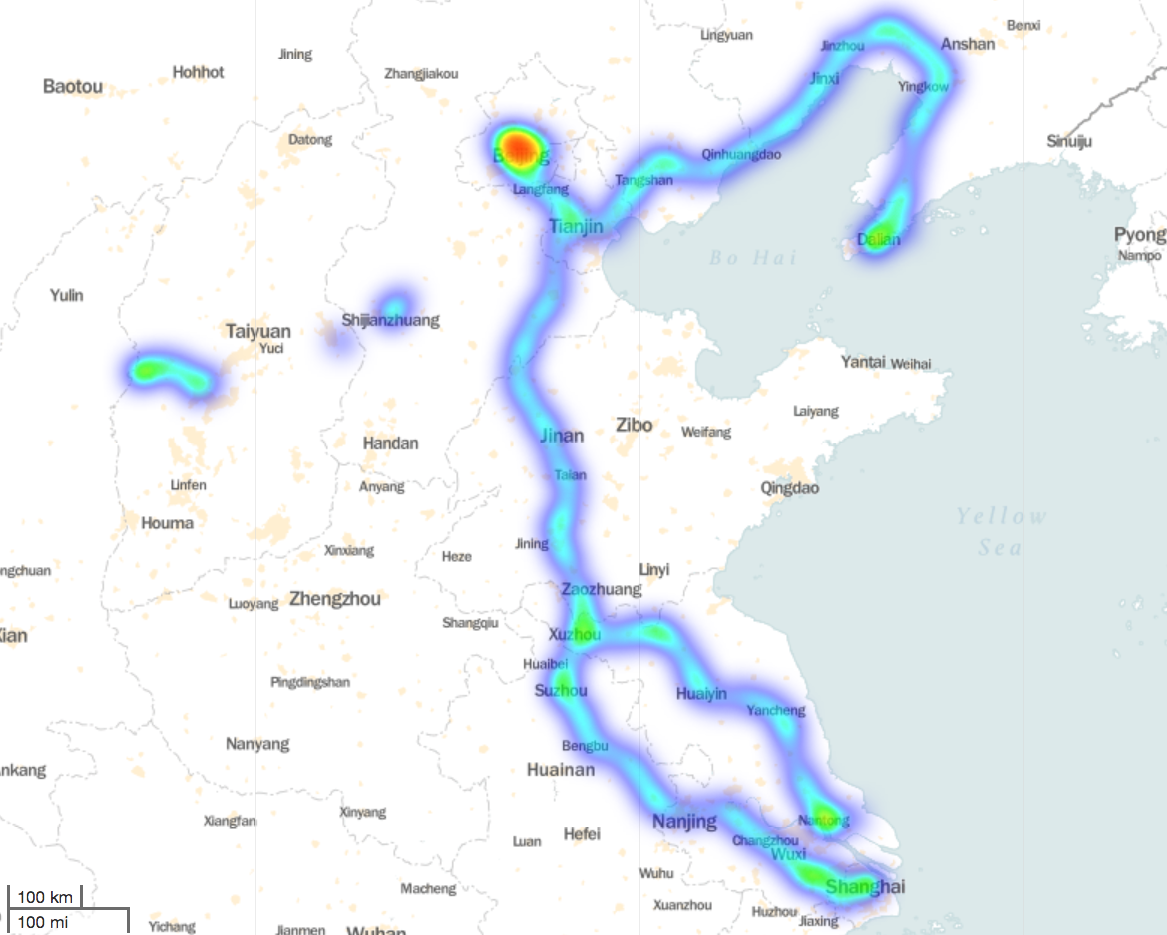}
    \caption{Location-Heatmap for the First 1000 Trajectories}
    \label{fig:route_heatmap}
\end{figure}

We also implemented a ClientManager which starts clients and assigns trajectories; the ClientManagers can be synchronized\footnote{Synchronizing ClientManagers is necessary when more than one machine is used as each ClientManager and its clients run on the same machine.} by providing a common start time.
To ensure determinism, we assigned an incrementing identifier to each trajectory in the dataset.
For example, if clients for the first 1000 trajectories should be started, plotting the trajectories' locations in a heatmap always yields the picture shown in figure~\ref{fig:route_heatmap}.

\subsection{GeoCheck Overhead} \label{subsec:geomatching_overhead}

With this experiment, we want to quantify the overhead of running the GeoChecks compared to solely running the ContentCheck.
We do this by reporting the operation throughput for subscription update operations (a data consumer updates its subscription) and subscription get operations (a data producer published a message and GeoBroker needs to determine the subscribers) when GeoChecks are enabled (GEO) and disabled (NoGEO).
These experiments were run with our subscription indexing structure only rather than using the complete GeoBroker implementation, as we do not want networking, message encoding/decoding, and other factors to influence our results.
The overhead highly depends on the implementation and type of indexing structure.
Therefore, this experiment primarily serves the purpose of getting a general understanding of parameters resulting in the highest GeoCheck overheads, proving that our implementation is efficient enough, and putting the results of the following use case evaluation into perspective.

Table~\ref{tbl:parameter_variations} shows our evaluated parameter set. The update/get ratio describes the ratio of update and get operations, e.g., (1/10) means that each update operation is followed by ten get operations.
\begin{table}[ht]
\centering
\begin{tabular}{lr} \toprule
Parameter                       & Set of Evaluated Values \\ \midrule
Number of Clients               & 1, 10, 100, 250, 500, 750, 1000 \\
Update/Get Ratio                & (99/1), (1/1), (1/10), (1/99) \\
GeoCheck                        & GEO and NoGEO \\
Granularity                     & 1, 10, 25, 50, 100 \\ \midrule
Total Number of Runs            & 56 \\ \bottomrule
\end{tabular}
\caption{Parameters of the GeoCheck Overhead Experiment}
\label{tbl:parameter_variations}
\end{table}
All experiments have been run on a single t3.xlarge AWS instance configured in unlimited mode for 15 minutes each.
During each experiment, the clients continuously send update or get messages in compliance with the update/get ratio to the subscription data structure.
Whenever they reach a location, they execute one operation; thus, we use the TeleportingClient to send requests as fast as possible.
In the NoGEO runs, clients update a single subscription to an example topic and get all subscriptions with a matching topic.
In the GEO runs, clients additionally supply a circular geofence around their current location with each subscription (radius = 0.01 degree which is roughly 1km at this latitude/longitude) and get only subscriptions with a matching topic around their current location (the message geofence also has a radius of 0.01 degree).

\begin{figure}[ht]
    \centering
    \includegraphics[width=\textwidth]{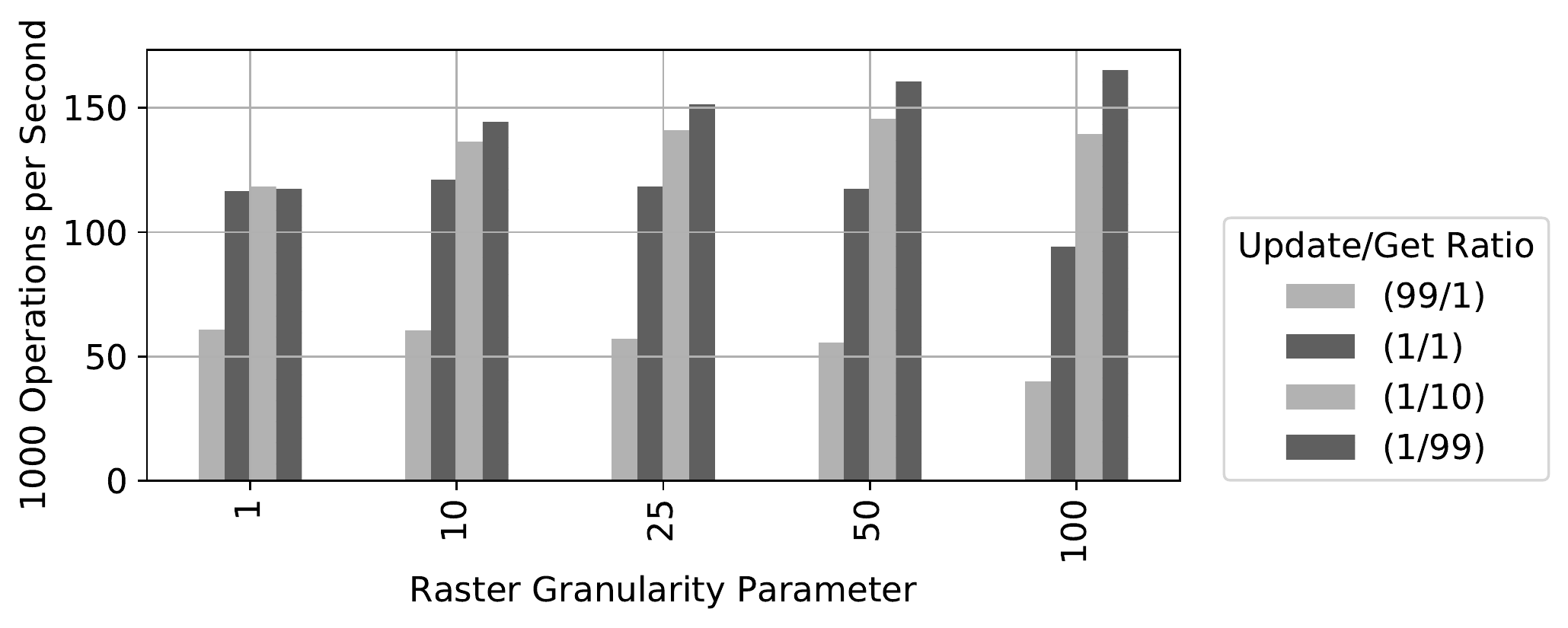}
    \caption{GEO Operation Throughputs for Different Granularity Values}
    \label{fig:storage_granularity}
\end{figure}

As explained in section~\ref{subsec:indexing_structure}, the granularity value can be used to tune subscription indexing structure performance (for the GEO runs); depending on the average geofence size of subscriptions, as well as the update/get ratio, different values yield the best result.
Figure~\ref{fig:storage_granularity} shows the number of operations that can be processed every second for 10 clients simultaneously.
When the raster granularity increases, the raster fields become smaller which improves get performance.
At the same time, this impairs update performance as each subscription must be added to more raster fields.
Note, that increasing the granularity value has the same effect on performance as reducing the geofence size, as the computational effort of algorithm~\ref{alg:raster_rectangle_algorithm} depends on the number of raster fields inside the bounding box.
Thus, we did not run additional experiments with different geofence sizes.

\begin{figure}[ht]
    \centering
    \includegraphics[width=\textwidth]{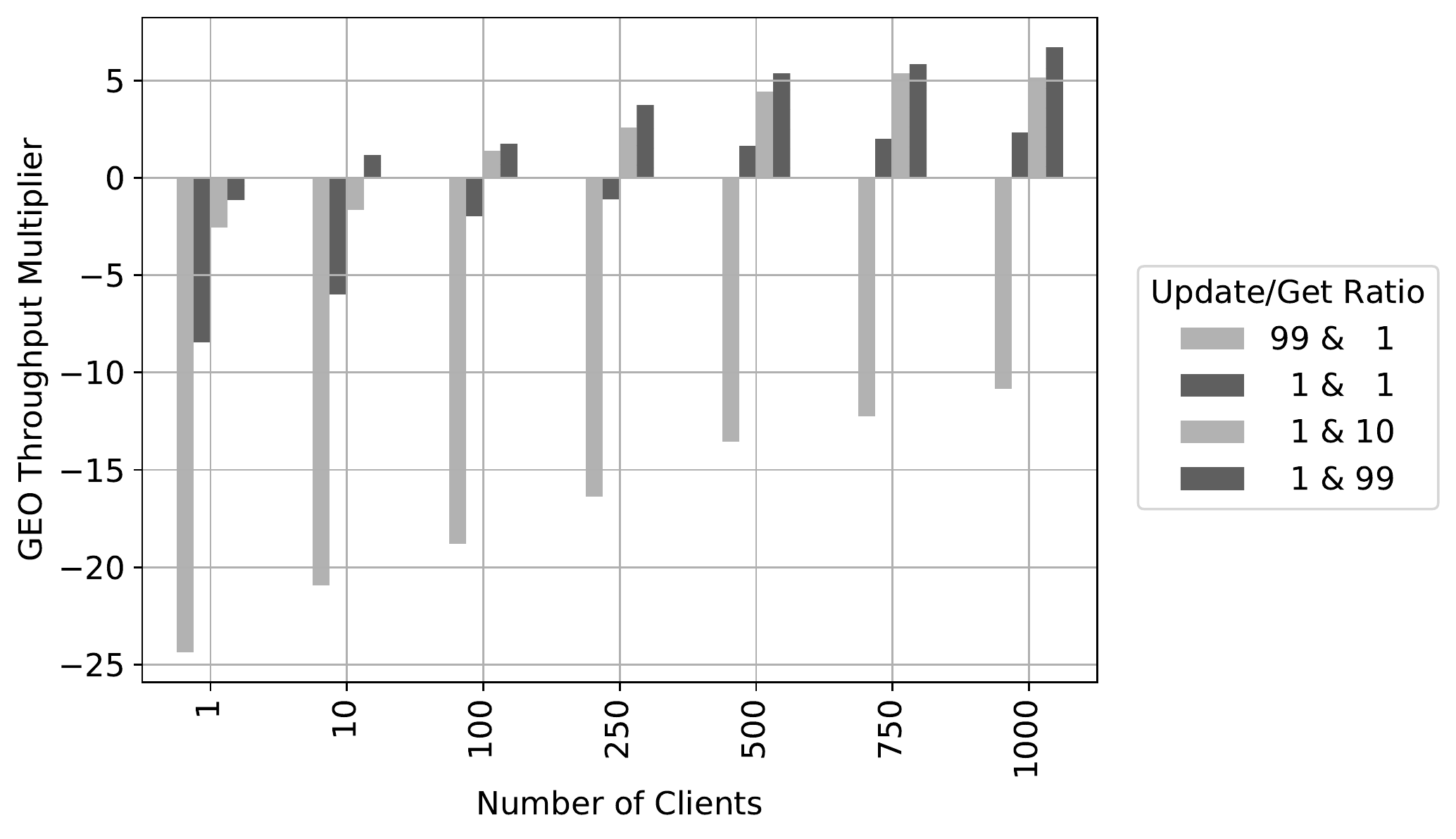}
    \caption{Using Geo-Context Information Increases the Performance of the Subscription Indexing Structure for Publish-Heavy Workloads}
    \label{fig:storage_normalVsNoContext}
\end{figure}

To compare GEO to NoGEO runs, we set the raster granularity to 25 as this value has a well balanced performance for all four update/get ratios (for the chosen geofence size).
Figure~\ref{fig:storage_normalVsNoContext} shows how using geo-contexts affects the performance of the subscription data structure.
Positive values mean that the GEO throughput is X times higher than the NoGEO throughput; negative values indicate the opposite.
If there is only a single client, the throughput is 24.37 times higher in the NoGEO run with the (99/1) update/get ratio, and 1.13 times higher for the (1/99) ratio, as updating the geofences in our storage component is more expensive than retrieving them.
However, this picture changes for higher client numbers, as NoGEO gets always return all existing subscriptions with matching topics while GEO only returns the subscriptions of nearby clients with a matching topic which leads to higher throughputs for publish-heavy workloads.
Thus, using geo-context information can help to significantly decrease the load on clients (irrelevant messages are simply not delivered), while also increasing the broker's performance.

Aside from the experiments above which evaluate GeoCheck overheads in a realistic scenario, we also determined the overhead of processing published messages when every message is always received by all clients (GEO and NoGEO).
For this, the ContentCheck and both GeoChecks must always be true for each connected client and published message, i.e., in GEO runs both geofences must comprise all clients.
While this is not realistic, as then checks could be omitted altogether, it allows us to determine an upper bound on the GeoCheck overhead.
We created an artificial workload; here, 100 clients publish a total of 11,547 messages over a period of 850 seconds.
We ran this workload on a single t3.large instance that comprises both clients and the broker.
To compare the performance between GEO and NoGEO, we measured the message delivery latency (MDL) for each individual message: MDL = $t_{received} - t_{send}$, with $t$ denoting the send and receive timestamp measured at each client.
The average MDL in the GEO run is 4.08\,ms, the average MDL in the NoGEO run is 3.79\,ms.
Thus, the overhead of processing published messages is about 7.7\% for this artificial workload.

\subsection{Application Use Case} \label{subsec:use_case}

With this experiment, we want to show how GeoBroker behaves in a realistic use case and analyze its performance in GEO and NoGEO runs, i.e., when using geo-context information is enabled and disabled.
In our use case, clients travel on their route and publish data to all other clients in close proximity when reaching a new location, so this application use case has some similarities to scenario 1 from~\cite{hasenburg_towards_2019}.
This data could be anything, e.g., surface condition information (roughness, surface, slipperiness), an image of the surroundings, broadcasted text messages, or requests for assistance.

In the GEO experiments, the consumer geofence ensures that only messages from nearby data producers are received, while the producer geofence ensures that data is not sent to data consumers outside a defined area, e.g., advertisement companies collecting user data.
For this, TravelClients connect to GeoBroker and execute the following three operations each time a location has been reached: First, send a ping message to update the current location.
Second, create/update the subscription to the topic \enquote{data} with a new circular consumer geofence (radius = 0.01 degree) around the current location.
Third, publish a new message to the topic \enquote{data} with a payload size of 750 byte and a producer geofence (radius = 0.01 degree) that also surrounds the current location.
Thus, each client acts as a data consumer and data producer at the same time.
In the NoGEO experiments, a subscription is created only once and thereafter messages are published without a geofence.
This means that clients do not have to update their subscriptions or locations. We also skip ping messages for NoGEO.

We ran each GEO and NoGEO experiment for 25 minutes with 250, 500, 750, and 1000 clients (250 clients are running together on one t3.xlarge instance).
We first ran all experiments with GeoBroker deployed on a t3.xlarge instance, and then repeated them with GeoBroker deployed on a t3.micro instance to study how GeoBroker copes with limited computational resources.
During the GEO experiments, update and get operations are alternating as we want the clients to update their location and publish a message each time they arrive at a location.
As for the corresponding (1/1) update/get ratio the subscription data structure has the highest throughput when the granularity is set to 10 (see figure~\ref{fig:storage_granularity}), we also set the granularity to 10 for this use case evaluation.

For each test run, we measured the publish latency which is the time between publishing a message and receiving an acknowledgment that the message has been sent to all data consumers that passed all three checks. As the message might still be on the wire, it may not yet have been received by all data consumers.
In general, however, the publish latency can be expected to be close enough to the delivery latency.
Furthermore, as this latency can be measured on the machine of the sending client, it is not affected by clock synchronization issues.

\begin{figure*}[ht]
   \subfloat[NoGEO Publish Latency]{\label{fig:NoGEO_750+1000}
      \includegraphics[width=.485\textwidth]{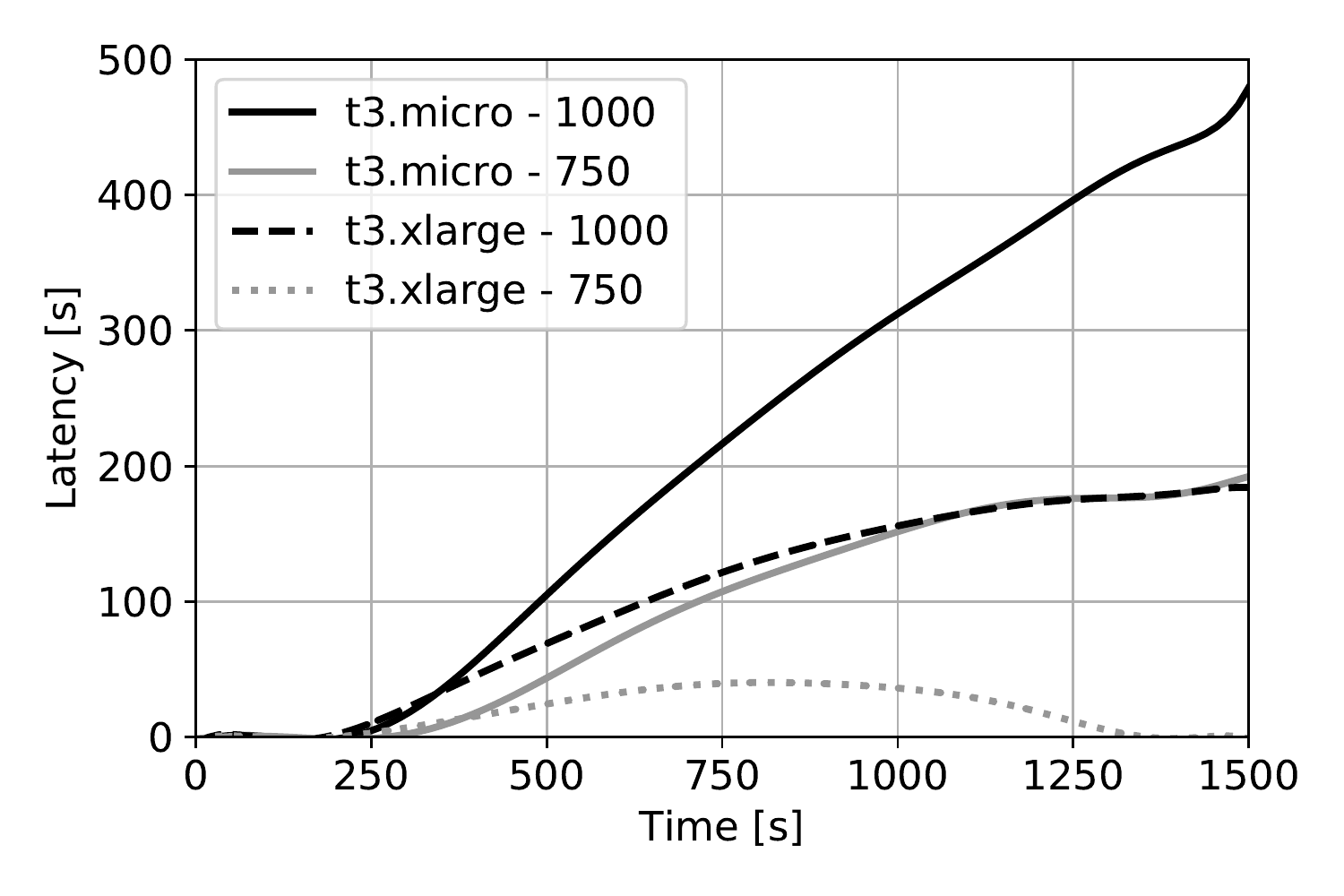}}
~
   \subfloat[Maximum GeoBroker CPU Load]{\label{fig:cpu}
      \includegraphics[width=.485\textwidth]{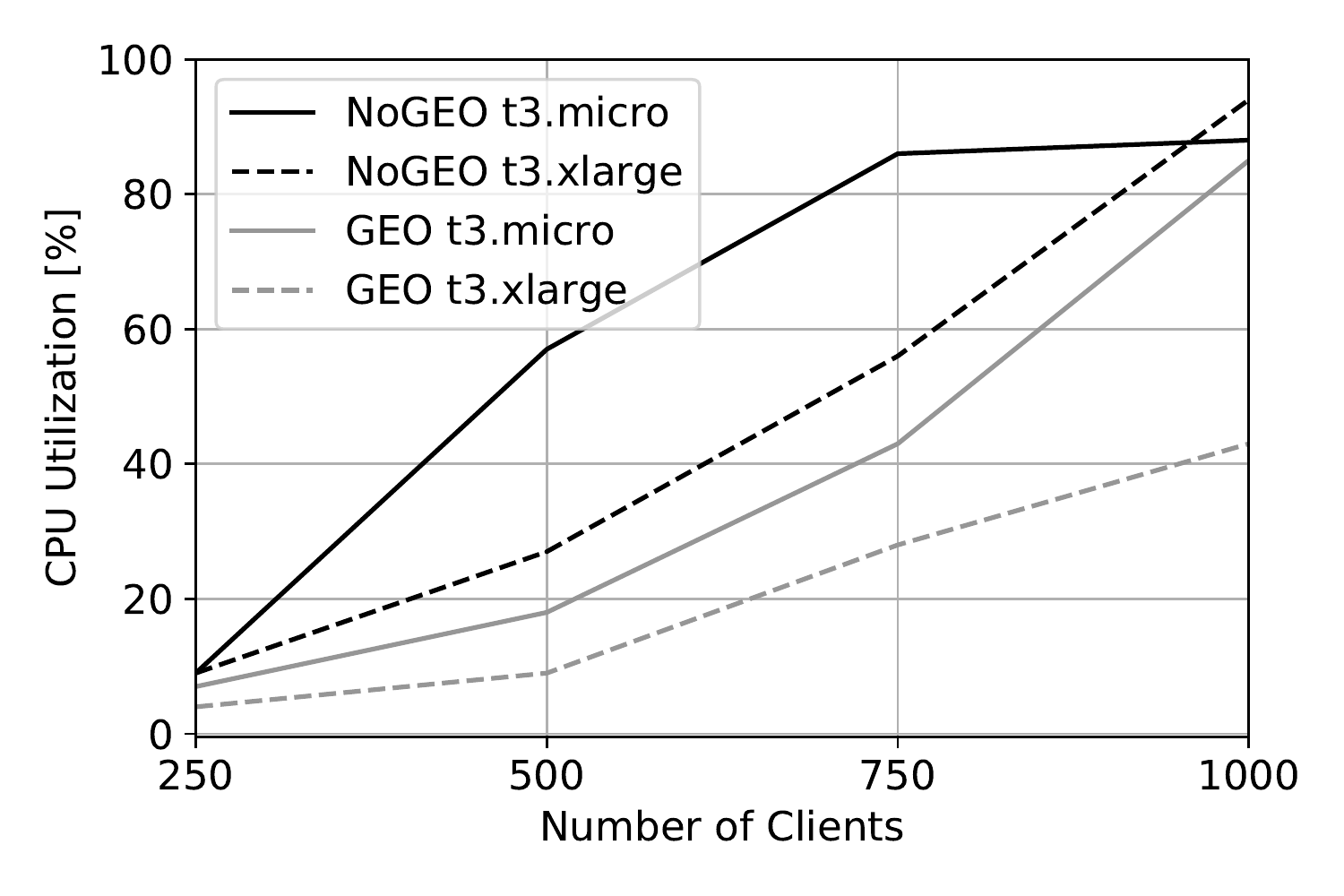}}

   \caption{Publish Latency and GeoBroker CPU Load in the NoGEO Experiment}
\end{figure*}

The first observation is that NoGEO only works for 250 and 500 clients (and for 750 clients on t3.xlarge).
For more clients, the latency continues to increase up to several minutes rather than stabilizing at a certain value; see also figure~\ref{fig:NoGEO_750+1000} which shows the publish latency for 750 and 1000 clients on t3.micro and t3.xlarge instances.
Furthermore, the message loss is substantial for these NoGEO runs, i.e., the t3.micro broker lost at least 22.5\% of its messages in 25 minutes with 750 clients, and 45.1\% of its messages with 1000 clients due to being overloaded.
These observations make sense when studying the number of messages that GeoBroker already delivers at 250 and 500 clients.
For 250 clients, the NoGEO broker delivers 9.4 million messages (GEO = 3.6 million messages).
This means, that for 250 clients, already more than 60\% of the transmitted messages have no relevance to receiving clients.
For 500 clients, the NoGEO broker delivers 39 million messages (GEO = 8.9 million messages).
When inspecting the CPU load (see figure~\ref{fig:cpu}), one can also identify that t3.micro NoGEO has a substantial amount of dropped messages including many connect and subscribe messages during the startup phase as the CPU load does not increase by a lot compared to 750 clients.

\begin{figure*}[ht]
    \center
   \subfloat[GEO]{\label{fig:GEO250-micro}
      \includegraphics[width=0.45\textwidth]{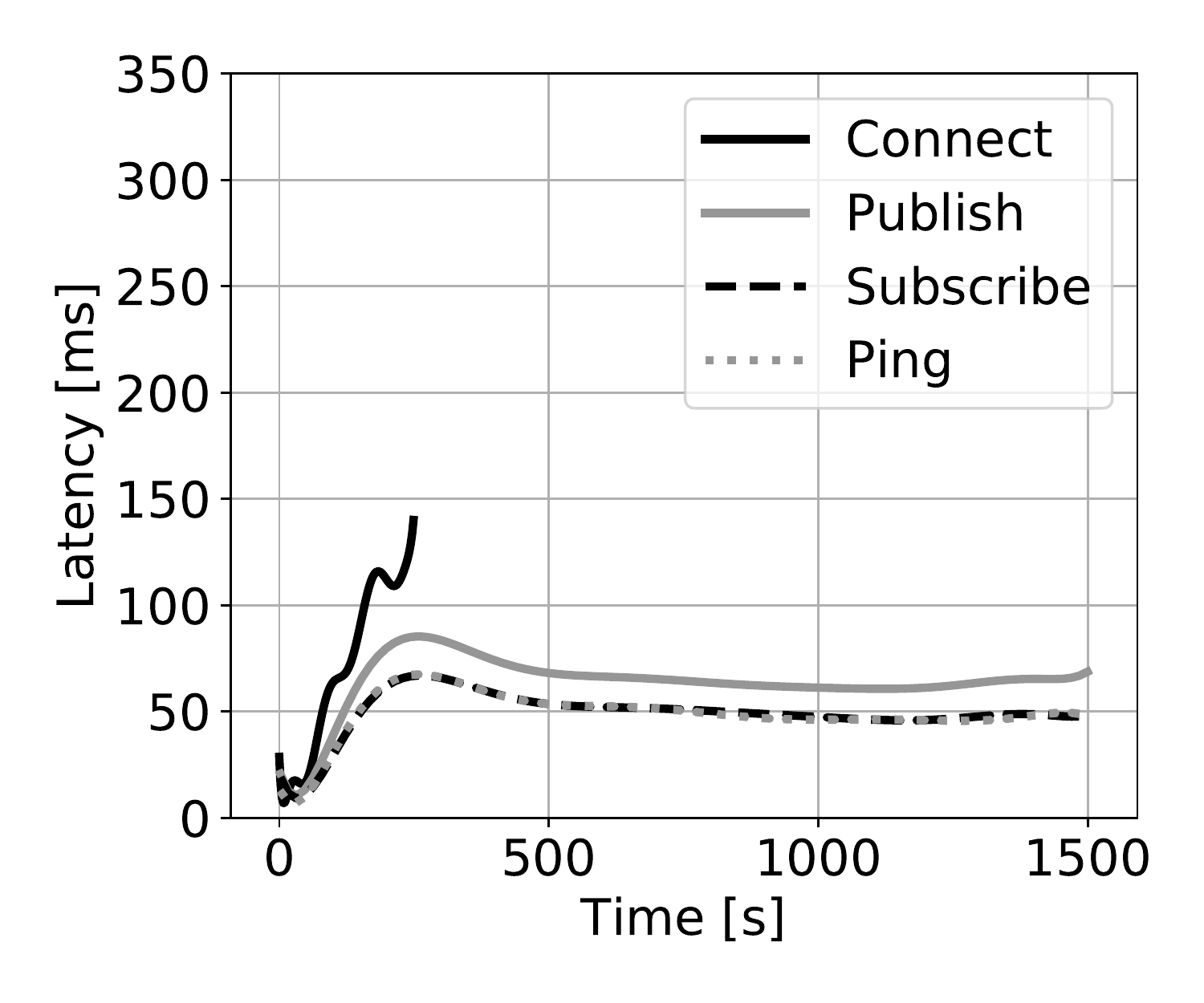}}
~
   \subfloat[NoGEO]{\label{fig:NoGEO250-micro}
      \includegraphics[width=0.45\textwidth]{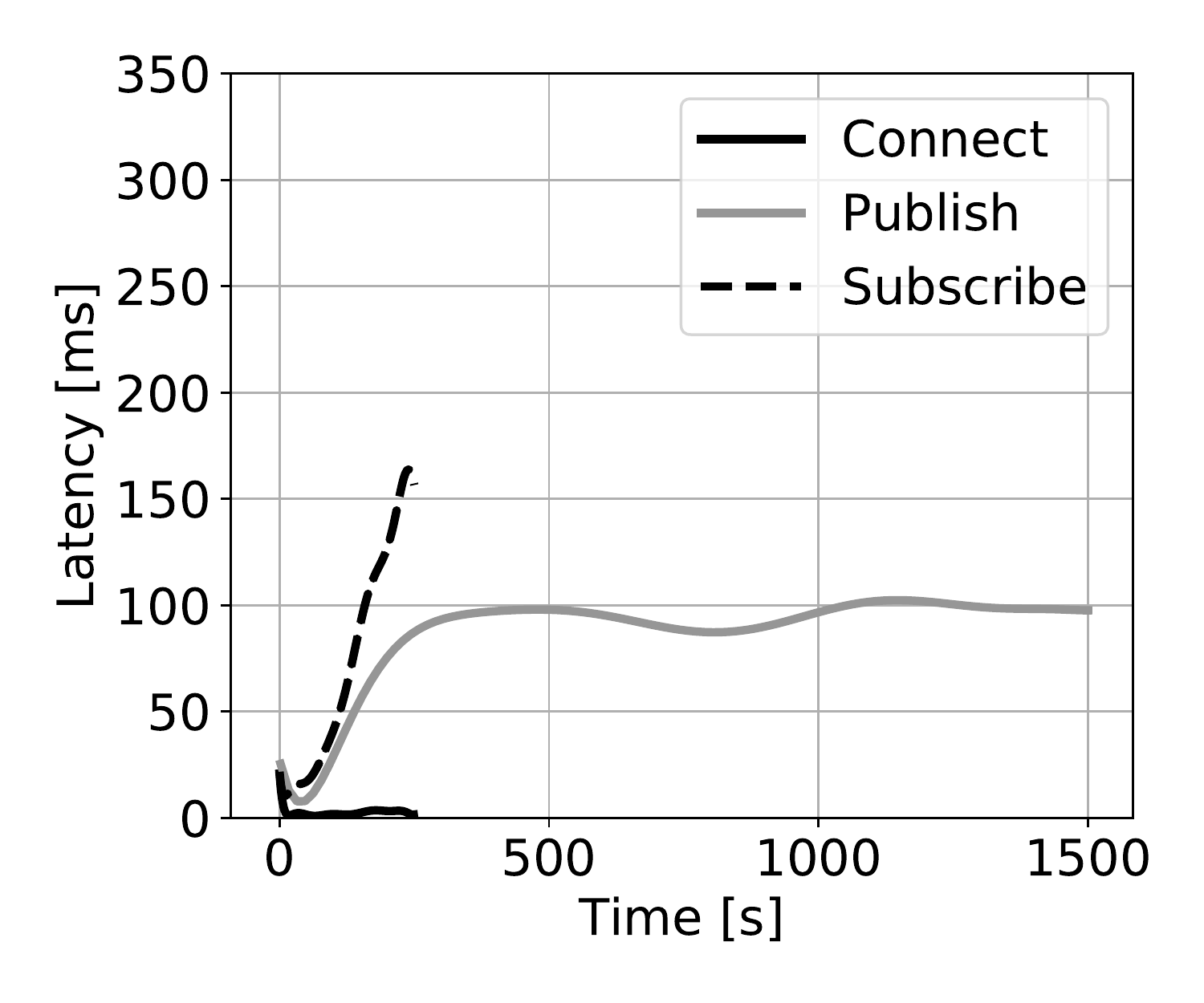}}

      \caption{Message Latency for 250 clients on t3.micro}
      \label{fig:250-micro}
\end{figure*}

\begin{figure*}[ht]
    \center
   \subfloat[GEO]{\label{fig:GEO500-micro}
      \includegraphics[width=0.45\textwidth]{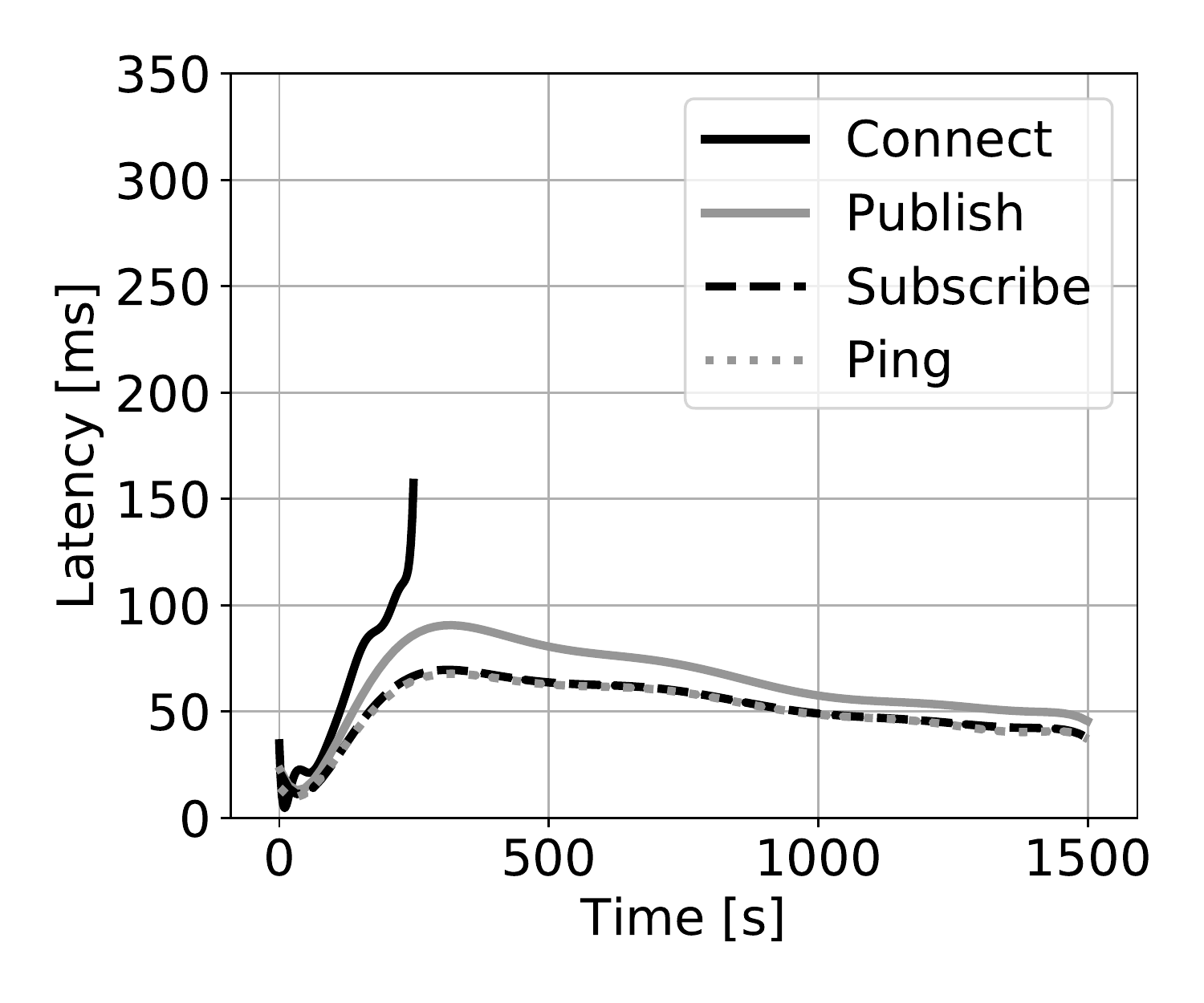}}
~
   \subfloat[NoGEO]{\label{fig:NoGEO500-micro}
      \includegraphics[width=0.45\textwidth]{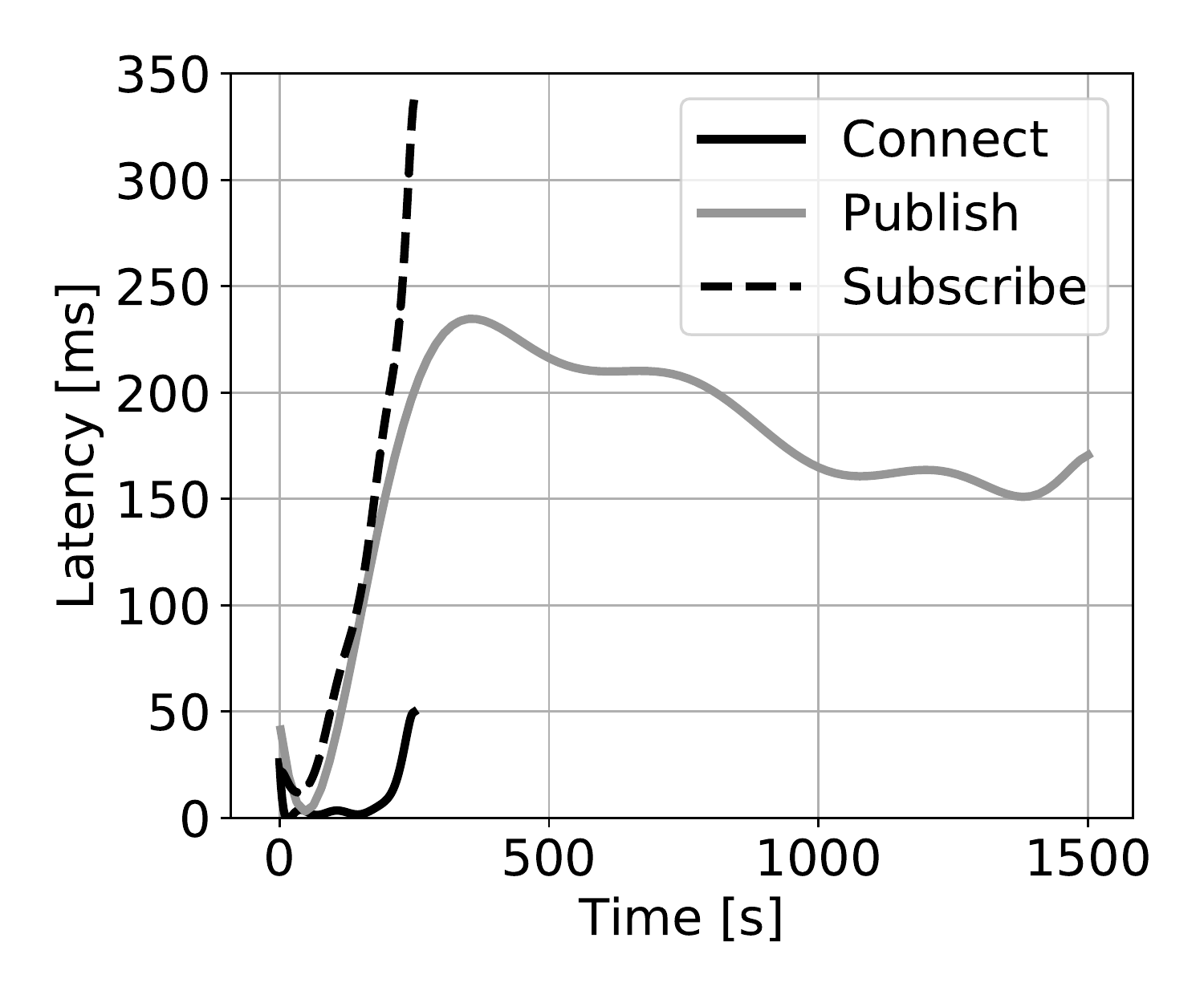}}

      \caption{Message Latency for 500 clients on t3.micro}
      \label{fig:500-micro}
\end{figure*}

In general, \textbf{GEO latency is always lower than NoGEO latency, even though additional computations are necessary}, the only exception is the 250 client experiment on the t3.xlarge instance.
Here, the NoGEO publish latency is on average 51ms, compared to 57ms for the GEO experiment.
Figure~\ref{fig:250-micro} and \ref{fig:500-micro} show the average publish latency over time for the GEO and NoGEO run with 250 and 500 clients on t3.micro\footnote{On t3.xlarge, the figures look very similar even though absolute latency values are smaller.}.

Besides the publish latency, the figures also contain the connect, ping, and subscribe latency (latency between sending a message and receiving an acknowledgment).
Note, that ping messages and subscription messages have an almost identical latency in the GEO run; NoGEO has no ping latency as no ping messages are sent.
Furthermore, the connect message latency stops after 250 seconds, as clients are started with a 1-second offset on each machine.

In general, the GEO runs have a good performance on the resource-constrained t3.micro instances, but when more than 750 clients are involved, the publish latency starts to rise quickly as well (it jumps from on average 101ms for 750 clients to about 293ms for 1000 clients).
However, for these large client numbers, GeoBroker can profit from stronger machines:
For 750 clients, the average publish latency is 101ms on t3.micro and 62ms on t3.xlarge.
Here, GeoBroker has to deliver 23.8 million messages.
For 1000 clients, the average publish latency is 293ms on t3.micro and 107ms on t3.xlarge.
Here, GeoBroker has to deliver 48.7 million messages.

Overall, this means that our prototype is sufficiently efficient and scales well vertically.
Furthermore, we showed that our approach can help to significantly reduce excess data dissemination for scenarios where geo-context matters; this preserves bandwidth and computational resources on data consumers and GeoBroker alike.

%% file: sections/07_discussion.tex

In this section, we discuss some of the design choices and limitations of GeoBroker (section~\ref{subsec:design_choices_and_limitations}).
Furthermore, we briefly describe open questions and new research opportunities (section~\ref{subsec:future_work}).

\subsection{Design Choices and Limitations} \label{subsec:design_choices_and_limitations}

The producer geofence can be used to limit data access without knowing the data consumers.
This can be a very useful feature in many situations.
Smart buildings, for example, can continuously publish their data to the same topic using a geofence that represents the building's shape to ensure that no one on the outside receives anything without having to worry about updating access control lists.
This, however, requires trust in the location provided by a data consumer.
While some solutions for that already exist~\cite{foamspace_corp._foam_2018}, these still have to prove their practical usability.

Doing GeoChecks as part of the message matching process is also not another form of content-based pub/sub.
In contrast to content-based filtering, GeoBroker also allows data producers to define criteria, i.e., a message might be filtered/not delivered to a data consumer based on restrictions put into place by data producers rather than data consumers only.
Furthermore, the geo-context of a client is not necessarily related to the content it receives/distributes, e.g., the location of a data producer is not related to to content of its published messages.
For instance, imagine a mobile ice cream vendor who continuously publishes the same data as content (i.e., the available ice cream flavors) while driving through a city. The city's ice cream aficionados will only be interested in the vendor's messages, when the ice cream vendor's position is within a geofence describing their area of interest.

Separating content and geo-context information also has the advantage of GeoBroker being payload-agnostic.
That is also one of the reasons why we chose not to encode location information in topics for the NoGEO experiments, as this would mix context and content information and force data producers and data consumers to agree on a topic structure (such as zip codes or geographic regions appended to each actual topic), even though they should be decoupled.

As of today, it remains unclear which type of indexing data structure is suited best for an application such as GeoBroker that leverages the four dimensions of geo-context.
For our prototype, we had very specific requirements such as being able to run the three message matching checks successively in a multi-threaded environment; production-ready performance has not been our goal.
While our subscription indexing structure is not the focus of this paper, we showed that it fulfills all these requirements and achieves a high performance.
Nevertheless, a more thorough evaluation is needed for a definitive conclusion which, however, is beyond the scope of this paper.

Finally, we want to emphasize that consumer and producer geofences can have arbitrary shapes.
However, if the shape gets more complex, the required \enquote{contain} operations become more computationally intensive which increases the get and update latency, as well as the CPU load of GeoBroker.
Nevertheless, as these checks are not carried out by the clients, this does not affect their performance.
Thus, the approach is still well suited for clients operating in constrained environments, but more broker resources might be required.
Our raster approach based on bounding boxes can alleviate parts of that extra complexity.

\subsection{Future Work} \label{subsec:future_work}

GeoBroker operation data, e.g., message flows and client interactions, could be used to extend approaches from the social networks field.
Existing work, e.g., \cite{doytsher_querying_2010,zheng_geolife:_2010,ferrari_extracting_2011,noulas_exploiting_2011,bao_location-based_2012}, often uses location traces collected from social media to identify correlation of users and locations, to derive recommendations, or to identify (emergency) events.
In addition, they often have to identify which physical area is affected by a certain message/tweet/picture/blog post---extracting this information is in many cases not trivial.
With GeoBroker, this information is available out of the box with high precision; e.g., a data producer already defines an area of relevance with the producer geofence.
Furthermore, with the consumer geofence, data consumers can precisely describe the areas they are interested in which is much more accurate than estimating this based on locations and information extracted from, for example, tweets.
Besides, GeoBroker can also be used to spread targeted information and emergency warnings that were identified through social networks data~\cite{longueville_omg_2009,xu_participatory_2016}, e.g., a forrest fire only affects people in a certain area and a related emergency warning must, therefore, only be distributed to people living in proximity to this area.

In its current version, GeoBroker is a single node broker.
However, we are already working on building a distributed version that supports sharding and replication for (local) clusters.
Using geo-context information for managing routing and replication in geo-distributed pub/sub broker deployments also appears to be a promising avenue to pursue.
With this, we plan to extend our work in the future to create a data distribution service that can be deployed at global scale.
Still, GeoBroker is already relevant in its current version for regional setups, e.g., when GeoBroker manages data distribution for a city-wide IoT deployment such as Santander~\cite{sanchez_smartsantander:_2014}.
In such a scenario, latency to the GeoBroker server is relatively low and the resulting load levels can easily be handled by a single machine as our experiments indicate.

%% file: sections/08_conclusion.tex

In this paper, we proposed to use the geo-contexts associated with IoT devices to control data distribution.
We showed that this can help to significantly reduce excess data dissemination for scenarios where geo-context matters while also facilitating the development of new (IoT) applications.
Our definition of geo-context comprises four dimensions: producer location, consumer location, producer geofence, and consumer geofence.
We discussed which of these four have been considered by related work.
In addition, we described the design of GeoBroker, a data distribution service leveraging geo-contexts of data consumers and data producers to control message distribution.

In our evaluation, we showed that ContentChecks combined with GeoChecks significantly reduce the number of transmitted messages for scenarios where geo-context matters, thus reducing the load on the data distribution service, the bandwidth consumption, and the number of messages that need to be processed by data consumers.
While this comes with the cost of additional computations that GeoBroker needs to run, we showed that this overhead is relatively small at low load levels and is, at higher load levels, more than offset by the performance improvements gained by only transmitting relevant messages.